\documentclass[ 
twocolumn,
showpacs,preprintnumbers,
bibnotes,
 amsmath,amssymb,
 aps,
 pra,
superscriptaddress,
longbibliography,
]{revtex4-1}

\usepackage[version=3]{mhchem} 
\usepackage[dvipsnames]{xcolor}
\usepackage{textcomp}
\usepackage{graphicx}
\usepackage{amsmath}
\usepackage{fixmath}
\usepackage{amsfonts}
\usepackage{amssymb}
\usepackage{braket}
\usepackage{subfigure}

\begin{document}

\title{Dissipation and spontaneous
emission in quantum electrodynamical density functional theory based on
optimized effective potential: A proof of concept study}

\author{A.~Kudlis}
\affiliation{%
 Faculty of Physics, ITMO University, St. Petersburg 197101, Russia}%
\author{I.~Iorsh}
\affiliation{%
 Faculty of Physics, ITMO University, St. Petersburg 197101, Russia}%
\author{I.~V.~Tokatly}
\affiliation{Nano-Bio Spectroscopy Group and European Theoretical Spectroscopy Facility (ETSF), Departamento de Polímeros y Materiales
Avanzados: Física, Química y Tecnología, Universidad del País Vasco, Avenida Tolosa 72, E-20018 San Sebastián, Spain}
\affiliation{IKERBASQUE, Basque Foundation for Science, 48009 Bilbao, Spain}
\affiliation{Donostia International Physics Center (DIPC), E-20018 Donostia-San Sebastián, Spain}
\affiliation{%
 Faculty of Physics, ITMO University, St. Petersburg 197101, Russia}%

\date{\today}

\begin{abstract}
We generalize the optimized effective potential (OEP) formalism in the quantum electrodynamical density functional theory (QEDFT) to the case of continuous distribution of photon modes, and study its applicability to dissipative dynamics of electron systems interacting with photons of lossy cavities. Specifically, we test whether this technique is capable of capturing the quantum features of electron-photon interaction related to spontaneous emission and the corresponding energy transfer from the electrons to cavity photons. For this purpose, we analyze a discrete three-site system with one electron coupled to photons of the cavity, which, in fact, is a minimal model allowing to eliminate classical radiation and the corresponding energy loss, but still have nontrivial density dynamics. By considering two typical spectral densities of photon modes, modeling (i) lossy cavity with Lorentzian broadening of photon peaks, and (ii) the Ohmic bath, and several representative dynamical regimes, we find that OEP-QEDFT demonstrates a good qualitative and quantitative performance, especially in the case when the disspation is dominated by one-photon processes.  
\end{abstract}

\maketitle
\section{Introduction}

Progress in the fields of cavity and circuit QED, and especially recent developments in polaritonic chemistry, also referred to as chemistry in cavity or QED-chemistry \cite{Hutchison2012,Ebbesen2016,Anoop2016,KowBenMuk2016,Zhong2017,Herrera2016,FliRivNar2018,FeiGalGar2018,ZhoMen2019,Hertzog2019,HerOwr2020} requires the development of theoretical methods for describing realistic many-electron systems strongly coupled to photons. The application of these methods ranges from the exploration of cavity-assisted phase transitions in many-electron systems~\cite{PhysRevLett.122.133602, PhysRevX.10.041027} to cavity engineering of the potential landscapes to taylor the photocatalysis~\cite{hutchison2012modifying}.  Such methods are expected to combine the accuracy of modern electronic structure theory with the ability to treat light fully quantum mechanically capturing the effects of strong light-matter interaction, typical for quantum optics \cite{Forn-Diaz2020,Kockum2019,Boite2020}.

Density functional theory (DFT) \cite{DreizlerGross1990} and its time-dependent counterpart (TDDFT) \cite{RunGro1984,TDDFT-2012,TDDFTbyUllrich} are the common methods of choice for modelling realistic materials because of their good balance between the accuracy and computational efficiency. It is therefore highly desirable to extend the DFT framework by including quantized electromagnetic degrees of freedom. Such QED generalization of the DFT concept, known as QED-TDDFT or QEDFT, has been indeed proposed few years ago \cite{Tokatly2013PRL,Ruggenthaler2014PRA}.  This theory being a reformulation of the many-body electron-photon problem treats photons on equal footing with electrons, and gives a formally exact access to the electron density and the electromagnetic field strength in the cavity. Different aspects of QEDFT have been studied in the last years \cite{Pellegrini2015PRL,Flick2015,Flick2017a,Flick2018,FliNar2018,Flick2019,Wang2021}. However, many general properties of this promising formalism remain poorly understood, while applications of QEDFT are still limited to the simplest level of mean field approximation.

Probably one of the most interesting features of QEDFT is that its structure allows for a natural inclusion of dissipative effects. As in many practically important situations a quantum system can not be considered perfectly isolated, the generalization of the TDDFT for modeling dissipative dynamics has always been a challenge. In the last two decades there were several proposals for including dissipation into TDDFT, based on master equation for density matrix \cite{BurCarGeb2005,Zhou2009,Zhou2010}, or starting from many-body stochastic Schr\"odinger equation \cite{VenAgo2007,AgoVen2008}. {It is worth noting that for a closed macroscopic system, dissipative effects related to internal excitation of the electron gas can be captured within the viscoelastic formulation of Vignale-Kohn current density functional \cite{VigUllCon1997,AgoVig2006}}. 

QEDFT is perfectly suited for quantum dissipative systems because it is formulated for electrons interacting with an arbitrary set of cavity modes. Without any modification of the formalism, the set of photon modes can be taken continuous with some spectral density and we get (TD)DFT for a system of electrons coupled to a quantum dissipative environment~\cite{Tokatly2013PRL}. Depending on a specific form of the spectral density, QEDFT may describe different physical systems ranging from molecules or nanostructures in realistic lossy cavities to many-electron systems coupled to the Caldeira-Leggett Ohmic bath \cite{CalLeg1983a,CalLeg1983b}. Despite a close relation of QEDFT to quantum dissipative systems was recognized essentially from its advent, this important aspect of the formalism remained practically unstudied till now. Very recently extensions of QEDFT to dissipative cavities with applications to the theory of the natural linewidth have been discussed \cite{Wang2021,Schaefer2021}, but only within the mean-field approximation for the electron-photon interaction. Similar to any TDDFT, in QED-TDDFT (QEDFT) dynamics of the electron density is mapped to the dynamics of fictitious nonintercating Kohn-Sham particles moving in the presence of an effective self-consistent potential which contains a mean-field (Hartree) and an exchange correlations (xc) contributions. The former corresponds to the classical coherent radiation \cite{Tokatly2013PRL,Pellegrini2015PRL} that describes the radiation reaction self-force \cite{Schaefer2021}, whereas the latter is responsible for all remaining purely quantum effects. In some situation, e.~g. in the linear response regime, the classical radiation reaction and the corresponding losses on coherent radiation can indeed dominate. However, by neglecting the xc potential in the mean-field approximation we completely ignore the quantum nature of the cavity filed and totally miss crucially important physical effects, such as spontaneous emission. For example, if in the course of dynamics the electronic subsystem preserves the inversion symmetry such that its center of mass is not moving, the coherent dipole radiation is absent and at the mean-field level the dynamics will be undamped, which is clearly unphysical. In reality the dissipation, that is, the energy transfer from the electrons to the cavity photons occurs via the spontaneous emission of incoherent radiation with zero expectation value of the field strength. In the QEDFT framework the physical behaviour should be restored by the quantum xc effects encoded in the xc potential. Apparently the potential doing this important job should be quite nontrivial, and it is absolutely unclear whether the existing approximations can do it, at least to some extent. This is the main question we address in this paper.

Specifically, we study the performance of the QED optimized effective potential (QED-OEP) approximation \cite{Pellegrini2015PRL} for lossy cavities and its ability to describe dissipation via spontaneous emission of incoherent radiation. To clearly disentangle the incoherent quantum radiation from the classical recoil effect we analyze the dynamical regimes where the classical radiation is absent and all dissipation is of purely quantum origin. Aiming at the proof of concept, we do this for a minimal 3-site tight-binding model in which a nontrivial density dynamics in the absence of the classical radiation reaction can be realized. By explicit numerical calculations we demonstrate that QED-OEP is able to capture the quantum dissipation both qualitatively, and to a very high accuracy quantitatively, at least in the regimes dominated by one-photon processes. 

The paper is orgnized as follows: In section II we provide the general description of the formalism used. In section III we  apply the formalism to the minimal lattice model, and discuss the properties of the exact solution as well as the Optimized Effective Potential (OEP) approximation. Section IV summarizes the main results of the numerical simulation, and Section V provides the conclusions and outlook.

\section{Statement of the problem: Quantum electrodynamical DFT for lossy cavities}

In this section, we describe the formalism used in the work, without resorting to a detailed description of a specific electronic subsystem. 

\subsection{The system Hamiltonian}
We start with the most general situation in which $N_{e}$ interacting electrons are coupled to $N_{\gamma}$ cavity modes. The position of $i$th electron is denoted by $\text{\bf{r}}_i$, while the canonical coordinate, momentum, and frequency of $\alpha$th photon mode are labeled by $q_{\alpha}$, $p_{\alpha}$, and $\omega_{\alpha}$ respectively. As usual in the context of cavity QED we assume that the size of the electronic system is much smaller that the wavelength of the relevant cavity modes and the electron-photon coupling is well described by the dipole approximation.  Keeping in mind the standard expression for the energy of transverse electromagnetic field $\int d{\bf r}\left[{\bf E}_{\perp}^{2}+{\bf B}^{2}\right]/8\pi$ as well as the connection of canonical variables with quantum amplitudes of electric displacement ($\hat{D}_{\alpha}=\sqrt{4\pi}\omega_{\alpha}\hat{q}_{\alpha}$) and magnetic field ($\hat{B}_{\alpha}=\sqrt{4\pi}\hat{p}_{\alpha}$), the general Hamiltonian of the electron-photon system within the Power–Zienau–Woolley (PZW) \cite{PowZie1959,Woolley1971} electric dipole gauge can be written as follows (see, for example, \cite{Tokatly2013PRL,Ruggenthaler2014PRA,Pellegrini2015PRL,AbeKhoTok2018EPJB}:
\begin{equation}\label{eqn:gen_ham}
\hat{H}=\hat{H}_e+\frac{1}{2}\sum_{\alpha=1}^{N_{\gamma}}\left[\hat{p}^2_{\alpha}+\omega^2_{\alpha}\left(\hat{q}_{\alpha}-\frac{\boldsymbol{\lambda}_{\alpha}}{\omega_{\alpha}}\hat{\text{\bf{R}}}\right)^2\right],
\end{equation}
where $\hat{\text{\bf{R}}}=\sum_{i=1}^{N_e}\textbf{r}_i$ is dipole moment operator of the electronic subsystem, and  the coupling constant $\boldsymbol{\lambda}_{\alpha}$ is determined by the electric field of the $\alpha$th mode at the location of the electronic system, $\boldsymbol{\lambda}_{\alpha}=\sqrt{4\pi}{\bf E}_{\alpha}$. The Hamiltonian of the electronic subsystem $\hat{H}_e$ consists of the kinetic energy $\hat{T}$, the Coulomb interaction $\hat{V}_{\text{C}}$, and the external potential $\hat{V}_{\textup{ext}}=\sum_{i=1}^{N_e} v_{\textup{ext}}(\textbf{r}_it)$, which is associated with an additional classical field applied to electrons. It is natural to rewrite the photon canonical variables in the second quantization formalism as follows,
\begin{eqnarray}
&&q_{\alpha}=\dfrac{1}{\sqrt{2\omega_{\alpha}}}\left(\hat{a}^{}_{\alpha}+\hat{a}^{\dagger}_{\alpha}\right),\quad
p_{\alpha}=-i\sqrt{\dfrac{\omega_{\alpha}}{2}}\left(\hat{a}^{\dagger}_{\alpha}-\hat{a}^{}_{\alpha}\right).\quad
\end{eqnarray}
In terms the operators $\hat{a}^{\dagger}_{\alpha}$ and $\hat{a}^{}_{\alpha}$ the part of Eq.~\eqref{eqn:gen_ham} responsible for the interaction can be divided into two contributions. The first, "cross term" reads as,
\begin{equation}
\label{eqn:Velph}
\hat{V}_{\textup{el-ph}}=\sum_{\alpha=1}^{N_{\gamma}}\sqrt{\frac{\omega_{\alpha}}{2}}(\hat{a}_{\alpha}+\hat{a}^{\dag}_{\alpha})\int d^3 \textbf{r}\left(\boldsymbol{\lambda}_{\alpha} \textbf{r}\right)\hat{n}(\textbf{r}),
\end{equation}
where $\hat{n}(\textbf{r})=\sum_{i=1}^{N_e}\delta(\textbf{r}-\textbf{r}_i)$ is the electron density operator. This is a typical fermion-boson coupling, which in particular generates an effective retarded interaction between electrons. The second contribution is the polarization energy of the electronic subsystem which can be written as  $\sum_{\alpha=1}^{N_{\gamma}}(\boldsymbol{\lambda}_{\alpha}\textbf{R})^2/2$ and has a form of an instantaneous electron-electron interaction. Thus, the electron-photon coupling induces an additional electron-electron interaction, 
\begin{eqnarray}
W_{\textup{ee}}(1,2)&=&\sum_{\alpha=1}^{N_{\gamma}}(\boldsymbol{\lambda}_{\alpha}\textbf{r}_1)(\boldsymbol{\lambda}_{\alpha}\textbf{r}_2)\mathcal{W}(t_1,t_2),\label{eqn:add_int_1}\\
\mathcal{W}(t_1,t_2)&=&\omega^2_{\alpha}D(t_1,t_2)+\delta(t_1-t_2),\label{eqn:add_int_2}
\end{eqnarray}
where the compact notation $1=(\textbf{r}_1t_1)$ is used. The first term in Eq.~\eqref{eqn:add_int_2} corresponds to the retarded photon-mediated interaction, where the photon propagator is determined in conventional manner: $\textup{i}D(t_1,t_2)\!\equiv\!\langle\mathcal{T}\,\{q_{\alpha}(t_1)\,q_{\alpha}(t_2)\}\rangle$. The second, instantaneous term in Eq.~\eqref{eqn:add_int_2} reflects the polarization energy in the electric part of the Hamiltonian \eqref{eqn:gen_ham}. 

Before we proceed further with QEDFT formalism, let us discuss the issue of the photon modes distribution.

\subsection{The distribution of modes}
In case of an ideal lossless cavity the photon modes are discrete and well defined, so that the main features of the light-matter interaction can be captured by considering only one or a few most relevant modes. In contrast, a realistic lossy cavity is characterized by a continuum of photon modes with a certain spectral density $\rho(\omega_{\alpha})$. Description of this situation within QEDFT formalism is the aim of the present paper. Of course in any practical numerical implementations the photon continuum is discretized, but the number of modes should be kept sufficiently large, $N_{\gamma}\gg 1$. Typically, to accurately mimic the dissipative effects one needs few thousands modes. In the following, as a reference situation, we also consider a single mode coupled to electron subsystem, which is formally introduced via a delta-type spectral density $\rho(\omega_{\alpha})=\delta_{\omega_r,\omega_{\alpha}}$, where $\omega_r$ is the resonance frequency. For lossy cavities we face with two typical situations. 

First, in order to take into account the experimentally observed broadening of spectral lines the delta-peaks of well defined discrete modes should be replaced by a smooth spectral density with a Lorentzian profile, 
\begin{eqnarray}\label{eqn:lor_prof}
\rho_{\textup{L}}(\omega_{\alpha},\gamma)&=& \Delta_{\omega}\dfrac{\gamma}{\gamma^2+(\omega_{\alpha}-\omega_r)^2},
\end{eqnarray}
where $\gamma$ is the loss rate defining the degree of broadening. The value of $\Delta_{\omega}$ is determined by the normalization condition $\sum_{\alpha=1}^{\textup{N}_{\gamma}}\rho_{\textup{L}}(\omega_{\alpha},\gamma)=1$ and distribution of spectral lines upon the formal discretization of the continuum. The simplest option is to take constant $\Delta_{\omega}$, indicating the uniform spacing between the modes. 

Second, the density of states can be flat imitating the Ohmic bath. In this case the distribution has the following form,
\begin{eqnarray}\label{eqn:unif_prof}
\rho_{\textup{O}}(\omega_{\alpha},\omega_{\textup{c}})&=&\dfrac{1}{N_{\gamma}}\theta(\omega_{\textup{c}}-\omega_{\alpha}),
\end{eqnarray}
where $\omega_{\textup{c}}$ is the cutoff frequency.

In both cases, the spectral density of states defines the distribution of the squared coupling constants \cite{Wang2021},  
\begin{eqnarray}
|\boldsymbol{\lambda}^{\textup{L}}_{\alpha}|^2=|\boldsymbol{\lambda}_{\textup{c}}|^2\rho_{\textup{L}}(\omega_{\alpha},\gamma),\quad
|\boldsymbol{\lambda}^{\textup{O}}_{\alpha}|^2=|\boldsymbol{\lambda}_{\textup{c}}|^2\rho_{\textup{O}}(\omega_{\alpha},\omega_{\textup{c}}).\quad \ 
\end{eqnarray}

\subsection{QEDFT formalism and optimized effective potential approximation}

The QED-(TD)DFT formalism is based on the statement that the many-body wave function of the combined electron-photon system $\Psi(\{\textbf{r}_j\},\{q_{\alpha}\},t)$ is uniquely determined by the electron density $n(\textbf{r}t)=\braket{\hat{\Psi}|\hat{n}|\hat{\Psi}}$ and the expectation values of the photon coordinate $q_{\alpha}(t)=\bra{\Psi}\hat{q}_{\alpha}\ket{\Psi}$. In order to compute the electron density, one can consider an auxiliary Kohn-Sham (KS) system of $N_e$ fictitious noninteracting particles, whose orbitals $\{\phi_j\}_{j=1}^{N_{e}}$ obey the following self-consistent equations:
\begin{eqnarray}\label{eq:KS_system}
\textup{i}\partial_t\phi_j(\textbf{r}t)=\left[-\dfrac{\nabla^2}{2}+v_s(\textbf{r}t)\right]\phi_j(\textbf{r}t),
\end{eqnarray}
with potential $v_{\textup{s}}=v_{\textup{ext}}+v_{\textup{eff}}$. The effective self-consistent potential $v_{\textup{eff}}$ consists of the mean-field contribution $v_{\textup{MF}}$ describing the classical radiation reaction and the exchange correlation (xc) potential $v_{\textup{xc}}$ which incorporates all the quantum many-body effects. The mean-filed contribution is expressed in terms of $n(\textbf{r}t)$ as follows \cite{Pellegrini2015PRL},
\begin{eqnarray}\label{eqn:v_mf}
v_\textup{MF}(\textbf{r}t)&&=\int d1\, W^{R}_{\textup{ee}}(\textbf{r}t,\textbf{r}_1t_1)\,n(\textbf{r}_1t_1),\nonumber\\
&&=\!\!\sum_{\alpha}(\boldsymbol{\lambda}_{\alpha}\textbf{r})\!\!\int_0^{t} \!\!\!dt_1\cos\!\left[\omega_{\alpha}(t\!-\!t_1)\right]\!(\boldsymbol{\lambda}_{\alpha}\dot{\textbf{R}}(t_1)),\quad
\end{eqnarray}
where $\textbf{R}(t)=\int d^3\textbf{r}\,\textbf{r}\, n(\textbf{r}t)$ is the expectation value of the dipole moment operator of electronic subsystem. The exchange potential, as in any DFT, is in general unknown and can be obtained only approximately. In this work, we resort to the generalization of the OEP approach proposed in Ref.~\onlinecite{Pellegrini2015PRL}. Below we briefly review its main points.

\begin{figure}[t]
    \centering
    \includegraphics[width=0.48\textwidth]{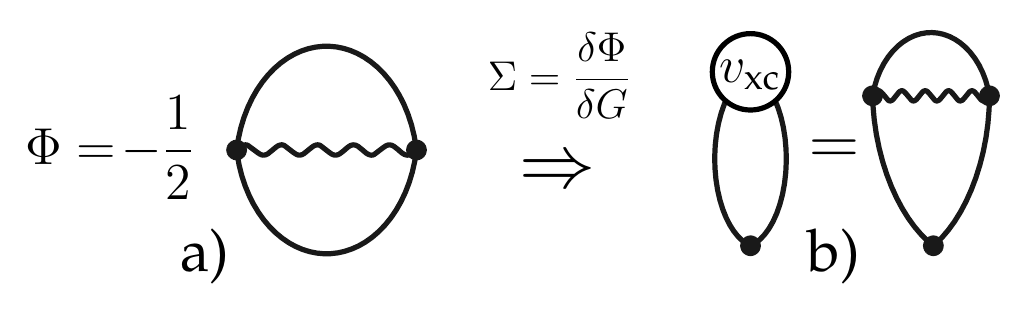}
    \caption{Diagrammatic representation of generating Baym functional (a), and the corresponding TDOEP equation (b) within lowest order conserving OEP approximation \cite{Pellegrini2015PRL}. Here solid lines stand for the KS Green functions $G_s$ and wiggled line is the cavity-induced interaction of Eq.~\eqref{eqn:add_int_1}.}    \label{fig:diag}
\end{figure}

The corresponding potential $v_{\textup{xc}}$ is the lowest or conserving OEP generated  by the Baym functional $\Phi$ shown in Fig.~\ref{fig:diag}(a) \cite{Tokatly2018PRB}.
The TDOEP equation for electron-photon system can also be understood as a linearized Sham-Schl\"uter equation~\cite{Leeuwen1996} on the Keldysh contour with the self-energy approximated by the one-photon exchange diagram. Diagrammatically this equation is presented in Fig.~\ref{fig:diag}(b), while analytically it reads,
\begin{eqnarray}\label{eqn:OEP}
&&\int d2 G_{\textup{s}}(1,2)v_{\textup{xc}}(2)G_{\textup{s}}(2,1)\nonumber\\
&&\qquad\qquad=\int d2 \int d3G_{\textup{s}}(1,2)\Sigma(2,3)G_{\textup{s}}(3,1),
\end{eqnarray}
where the electron self-energy $\Sigma$ is defined as follows,
\begin{equation}
    \Sigma(1,2)=iG_{\textup{s}}(1,2)W_{\textup{ee}}(2,1),
\end{equation}
with the free photon propagator of Eq.~\eqref{eqn:add_int_1}. It is worth reminding that the general idea of deriving conserving approximations in TDDFT based on the Baym $\Phi$-functional \cite{Baym1962} has been proposed in Ref.~\onlinecite{Barth2005}. Recently it has been adopted to QEDFT \cite{Tokatly2018PRB}. The explicit TDOEP equation in terms of KS orbitals reads as follows,
\begin{eqnarray}\label{eqn:TDOEP}
&&\textup{i}\sum_{i,j}\!\int_{-\infty}^{t}\!\!\!\!\!\!dt_1\!\left[\bra{\phi_i(t_1)}v_{\textup{xc}}(t_1)\ket{\phi_j(t_1)}f_i-S_{ij}(t_1)\right]\phi^*_j(t)\phi_i(t)\nonumber\\
&&+c.c.=0,
\end{eqnarray}
Here functions $S_{ij}(t_1)$ are defined as:
\begin{eqnarray}\label{eqn:s_matrix}
&&S_{ij}(t_1)\!=\!\sum_{k,\alpha}\int_{-\infty}^{t_1}\!\!\!\!\!\!dt_2\, d^{\alpha}_{ik}(t_2)d^{\alpha}_{kj}(t_1)[(1-f_i)f_k \mathcal{W}^>(t_1,t_2)\nonumber\\
&&\qquad\qquad-f_i(1-f_k)\mathcal{W}^<(t_1,t_2)],
\end{eqnarray}
where $f_i$ are occupation numbers of KS orbitals, $d^{\alpha}_{ik}(t)=\boldsymbol{\lambda}_{\alpha}\bra{\phi_i(t)}\textbf{r}\ket{\phi_k(t)}$ is the dipole matrix element projected on the coupling constant of the $\alpha$-mode, and photon propagators $\mathcal{W}^\gtrless(t_1,t_2)$ are expressed as:
\begin{eqnarray}
\mathcal{W}^\gtrless(t_1,t_2)=-\omega^2_{\alpha}\Big(\frac{\textup{i}}{2\omega_{\alpha}}\Big)e^{\pm \textup{i}\omega_{\alpha}(t_2-t_1)}\pm \delta(t_1-t_2).\qquad \label{eq:photon_propag}
\end{eqnarray}
The functions $S_{ij}$, in fact, represent the matrix elements of the self-energy, consisting of various combinations of different electronic states and photon propagators from Eq.~\eqref{eq:photon_propag} which describe the absorption and emission processes. The orbitals $\{\phi_j\}_{j=1}^{N_{e}}$ are the solution of system~\eqref{eq:KS_system} with stationary initial conditions: $\phi_j(\textbf{r}t)=\phi_j(\textbf{r})e^{-\textup{i}\varepsilon_jt}$ for negative times, where $\varepsilon_j$ are eigenvalues of the corresponding stationary problem: $\varepsilon_j\phi_j(\textbf{r})=\left[-\nabla^2/2+v_s(\textbf{r})\right]\phi_j(\textbf{r})$.

In this work, we focus on the QEDFT description of spontaneous radiation which is of a purely quantum-mechanical nature. The classical coherent electromagnetic radiation of a moving charge enters QEDFT formalism via the mean-field potential of Eq.~\eqref{eqn:v_mf} that is, in fact, a radiation reaction potential. In the QEDFT context the dissipation effects associated to mean-field radiation reaction has been considered recently \cite{Wang2021,Schaefer2021}. In contrast, a quantum spontaneous radiation and the corresponding dissipation, which is a purely xc effect encoded in $v_{\textup{xc}}$, has never been analyzed. In the following we completely suppress the classical dipole radiation by choosing the external potential with a certain symmetry. In other words, to separate the quantum radiation effects we consider the regimes of dynamics with the mean-field potential identically equal to zero, which leads to equality of effective and xc potentials $v_{\textup{eff}}=v_{\textup{xc}}$.
 
\section{Minimal lattice model for studying quantum dissipation}

\begin{figure}[t]
    \centering
    \includegraphics[width=0.49\textwidth]{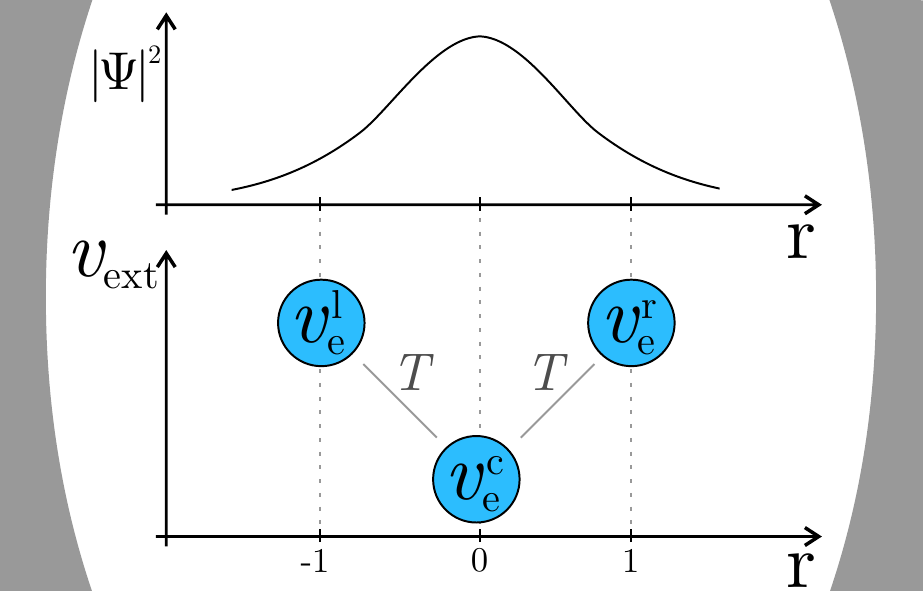}
    \caption{Schematic representation of three-site lattice in a cavity. The tunneling of electron is allowed only through the central site with hopping rate $T$. In addition to interaction of electron with cavity photons, it experiences the on-site external potential $v_{\textup{e}}^{\textup{i}}$.}
    \label{fig:system}
\end{figure}

The minimal model required for our purposes -- to demonstrate spontaneous emission -- is a discrete three-site tight-binding model with one electron coupled to cavity modes. This system is schematically shown in Fig.~\ref{fig:system}. We choose a reflection symmetric external potential that always produces a symmetric distribution of the electron density with a time independent (zero) dipole moment. As a result, it is possible to have nontrivial density dynamics with identically vanishing mean-field potential.  

In the next two subsections, we describe some technical details of solving the OEP-QEDFT problem for our model, as well as its numerically exact solution. The latter is used as a benchmark to assess the quality of the OEP approximation for lossy cavities with a continuum of photon modes.

\subsection{Exact solution}

For one electron on three sites interacting with one photon mode the Hamiltonian~\eqref{eqn:gen_ham} simplifies as follows,
\begin{eqnarray}\label{eqn:latt_ham}
&&H=-\hat{T}+\hat{V}_{\textup{ext}}+\dfrac{\lambda^2_{\textup{c}}}{2}\hat{R}^2\nonumber\\
&&\qquad\qquad+\omega\Big(\hat{a}^{\dagger}\hat{a}+\dfrac{1}{2}\Big)+\hat{R}\sqrt{\dfrac{\omega}{2}}\lambda_{\textup{c}}\Big(\hat{a}^{\dagger}+\hat{a}^{}\Big),\qquad
\end{eqnarray} 
The matrices $\hat{T}$, $\hat{R}$, and $\hat{V}_{\textup{ext}}$ represent in the tight-binding basis the kinetic energy operator, the operator of the dipole moment, and the external potential operator, respectively, 
\begin{eqnarray} 
&&\hat{T}=\begin{bmatrix}
0 & T & 0 \\
T & 0 & T \\
0 & T & 0 
\end{bmatrix}, \ \hat{R}=\begin{bmatrix}
r_\textup{r} & 0 & 0 \\
0 & r_\textup{c} & 0 \\
0 & 0 & r_\textup{l} 
\end{bmatrix}, \ \hat{V}_{\textup{ext}}=\begin{bmatrix}
v_\textup{e}^\textup{l} & 0 & 0 \\
0 & v_\textup{e}^\textup{c} & 0 \\
0 & 0 & v_\textup{e}^\textup{r} 
\end{bmatrix},\nonumber
\end{eqnarray}
where $T$ is the hopping rate, and $v_\textup{l}$, $v_\textup{c}$, and $v_\textup{r}$ are components of an external potential on the left, central, and right sites, respectively. In general, the onsite potentials can be time-dependent. For convenience, we choose the origin of coordinates at the central site, and assume the following values for the site coordinates: $r_\textup{l}=-1$, $r_\textup{c}=0$, and $r_\textup{r}=1$. The wave function of total system can be written as $\Psi_{q}(t)=(\psi^\textup{l}_q(t),\psi^\textup{c}_q(t),\psi^\textup{r}_q(t))$, where the  real  continuum  variable $q$ denotes the photonic canonical coordinate. In terms of these functions, the time-dependent Schr\"odinger equation governing  the  time  evolution  of  the  electron-photon state from a given initial one reads as follows,
\begingroup
\allowdisplaybreaks
\begin{eqnarray}
&&\textup{i}\partial_t\psi^\textup{l}_q(t)=-T\psi^\textup{c}_q(t)
\nonumber\\
&&+\Bigg[v_\textup{l}+\frac{\lambda^2_{\textup{c}}}{2}+\omega\Big(\hat{a}^{\dagger}_{}\hat{a}^{}_{}+1/2\Big)-\sqrt{\frac{\omega}{2}}\lambda_{\textup{c}}\Big(\hat{a}^{\dagger}+\hat{a}^{}\Big)\Bigg]\psi^\textup{l}_q(t),\quad \ \ \label{eq:wave_fun_l}\\
&&\textup{i}\partial_t\psi^\textup{r}_q(t)=-T\psi^\textup{c}_q(t)
\nonumber\\
&&+\Bigg[v_\textup{r}+\frac{\lambda_{\textup{c}}^2}{2}+\omega\Big(\hat{a}^{\dagger}_{}\hat{a}^{}_{}+1/2\Big)-\sqrt{\frac{\omega}{2}}\lambda_{\textup{c}}\Big(\hat{a}^{\dagger}+\hat{a}^{}\Big)\Bigg]\psi^\textup{r}_q(t),\quad \ \  \label{eq:wave_fun_r}\\
&&\textup{i}\partial_t\psi^\textup{c}_q(t)=-T\psi^\textup{l}_q(t)-T\psi^\textup{r}_q(t)
\nonumber\\
&&+\Bigg[v_c+\omega\Big(\hat{a}^{\dagger}_{}\hat{a}^{}_{}+1/2\Big)\Bigg]\psi^\textup{c}_q(t).\quad \label{eq:wave_fun_c} \ \ 
\end{eqnarray}
\endgroup

\noindent Due to the gauge invariance the physics should not be changed if we modify the potential by adding a global time-dependent quantity. For this reason we are free to assume $v_{\textup{e}}^\textup{c}=-v_{\textup{e}}^\textup{l}-v_{\textup{e}}^\textup{r}$. In the single-mode case, for all the regimes considered in this paper, the system of equations~\eqref{eq:wave_fun_l}-\eqref{eq:wave_fun_c} is solved numerically by the proper truncation of the photon Fock space. Specifically, in our calculations the convergence of the results is typically achieved for the Fock space dimension not exceeding $100$.

In order to take into account dissipation in a lossy cavity, one should add coupling to $N_{\gamma}\gg 1$ modes representing the photon continuum with the spectral density $\rho(\omega)$. This is introduced via the following replacements of the photon-dependent terms in the Hamiltonian~\eqref{eqn:latt_ham},   
\begin{eqnarray}
&&\sqrt{\dfrac{\omega}{2}}\lambda_{\textup{c}}\Big(\hat{a}^{\dagger}+\hat{a}^{}\Big)\rightarrow \sum\limits_{\alpha=1}^{N_{\gamma}}\sqrt{\dfrac{\omega_{\alpha}\rho(\omega_{\alpha})}{2}}\lambda_{\textup{c}}\Big(\hat{a}^{\dagger}_{\alpha}+\hat{a}_{\alpha}^{}\Big),\quad\\
&&\omega\Big(\hat{a}^{\dagger}\hat{a}+\dfrac{1}{2}\Big)\rightarrow \sum\limits_{\alpha=1}^{N_{\gamma}}\omega_{\alpha}\hat{a}^{\dagger}_{\alpha}\hat{a}_{\alpha},\quad
\end{eqnarray}
where the irrelevant vacuum energy is omitted. For dissipative dynamics in the direct solution of the electron-photon problem we limit our consideration only by one-photon states. This dramatically simplifies computations, and is, in fact, sufficient in the coupling range we consider here. On the other hand, it makes more natural a comparison with the lowest order OEP approximation based on the one-photon exchange diagram, see Fig.~\ref{fig:diag}.

Let us now define the quantities which should be studied. Obviously, the on-site densities ($n_\textup{l}(t)$, $n_\textup{c}(t)$, $n_\textup{r}(t)$), being the basic variables of any DFT, are of primary importance. Within our one-electron lattice model~\eqref{eqn:latt_ham}, these densities can be computed as follows,
\begin{equation}
    n_{i}(t)=\int|\psi^{i}_q(t)|^2 dq, \quad i=\textup{l},\textup{c},\textup{r}.
\end{equation}
All results, however, will be given for the following composite quantity,
\begin{equation}\label{eqn:delta_n}
\Delta n (t)= n_{\textup{r}}(t)+n_{\textup{l}}(t)-n_{\textup{c}}(t),
\end{equation}
which is the difference between the occupations of the side sites and the central site. Due to the chosen symmetry and the conservation of the number of particles, the value of $\Delta n (t)$ gives a complete picture of electron dynamics on the three-site lattice. At this point, we recall that the exact solution of the model~\eqref{eqn:latt_ham} serves only as benchmark to measure the accuracy of the results obtained within OEP approximation. Unfortunately, the on-site density is a notably rough variable, which is not always sufficient for the adequate comparison of the system behaviour obtained from the exact and approximate solutions. In this regard, the analysis of the xc potential in the KS Hamiltonian, which is another key object of DFT, can help to discriminate the results obtained by different methods. To reconstruct the xc potential from the exact solution we adopt the inversion procedure used to prove the mapping theorems for the lattice TDDFT \cite{FarTok2012PRB} and lattice QEDFT \cite{FarTok2014PRB}. 

In our case the KS Hamiltonian reads,
\begin{eqnarray} \label{eqn:ks_syst}
&&\hat{H}_{\textup{KS}}=-\hat{T}+\hat{V}_{\textup{S}}=\begin{bmatrix}
v_\textup{s}^\textup{l} & -T & 0 \\
-T & v_\textup{s}^\textup{c} & -T \\
0 & -T & v_\textup{s}^\textup{r} 
\end{bmatrix},
\end{eqnarray}
where KS potential $v_{\textup{s}}$ is related to the xc potential $v_{\textup{xc}}$ as $v_{\text{s}}^{i}=v_{\text{eff}}^{i}+v_{\textup{e}}^{i}=v_{\textup{xc}}^{i}+v_{\textup{e}}^{i}$. For a given density, the KS potential can be obtained by solving self-consistently the KS equation together with the following system of algebraic equations,
\begin{equation}\label{eqn:invert_problem}
\hat{K}[\Psi_{\textup{S}}]V_{\textup{S}}= S[\ddot n,\Psi_{\textup{S}}],
\end{equation}
Here $\Psi_{\textup{S}}$ is a vector-solution of the time-dependent KS equation with the Hamiltonian~\eqref{eqn:ks_syst}, $\hat{K}$ is a real symmetric $3\times 3$ matrix with elements 
\begin{equation}\label{eqn:}
k_{i,j}[\Psi_{\textup{S}}]=2\,\textup{Re}\left[T_{i,j}\rho_{i,j}-\delta_{i,j}\sum_{n}T_{i,n}\rho_{i,n}\right],
\end{equation}
$V_{\textup{S}}$ is a three-dimensional vector composed of on-site KS potentials $v_{\text{s}}^{j}$, and $S$ is a vector with components
\begin{eqnarray}
&&s_{j}[\ddot n,\Psi_{\text{S}}]=- \ddot n_{j} -q_{j}[\Psi_{\textup{S}}].
\end{eqnarray}
In the above equations, $q_j$ and $\rho_{i,j}$ are defined as follows,
\begin{eqnarray}
&&q_{j}=-2\,\textup{Re}\sum\limits_{i,n} T_{j,i}\Big[T_{i,n}\rho_{j,n}-T_{j,n}\rho_{i,n}\Big],\\
&&\rho_{i,j}=\Psi_{\textup{S},i}^*\Psi_{\textup{S},j}, \quad  T_{j,n} = (\hat{T})_{j,n}
\end{eqnarray}
Values $\ddot n_j(t)$ of second time derivatives of the densities enter the problem as an input taken from the exact solution of the system~\eqref{eq:wave_fun_l}-\eqref{eq:wave_fun_c}. The existence of the unique solution to the described reconstruction problem has been demonstrated in Ref.~\cite{FarTok2012PRB} for the purely electronic lattice TDDFT and for its QED generalization~\cite{FarTok2014PRB}. 

\subsection{OEP approximation in QEDFT}
\begin{figure*}[t!]
\subfigure{\label{fig:1}\includegraphics[width=0.49\textwidth]{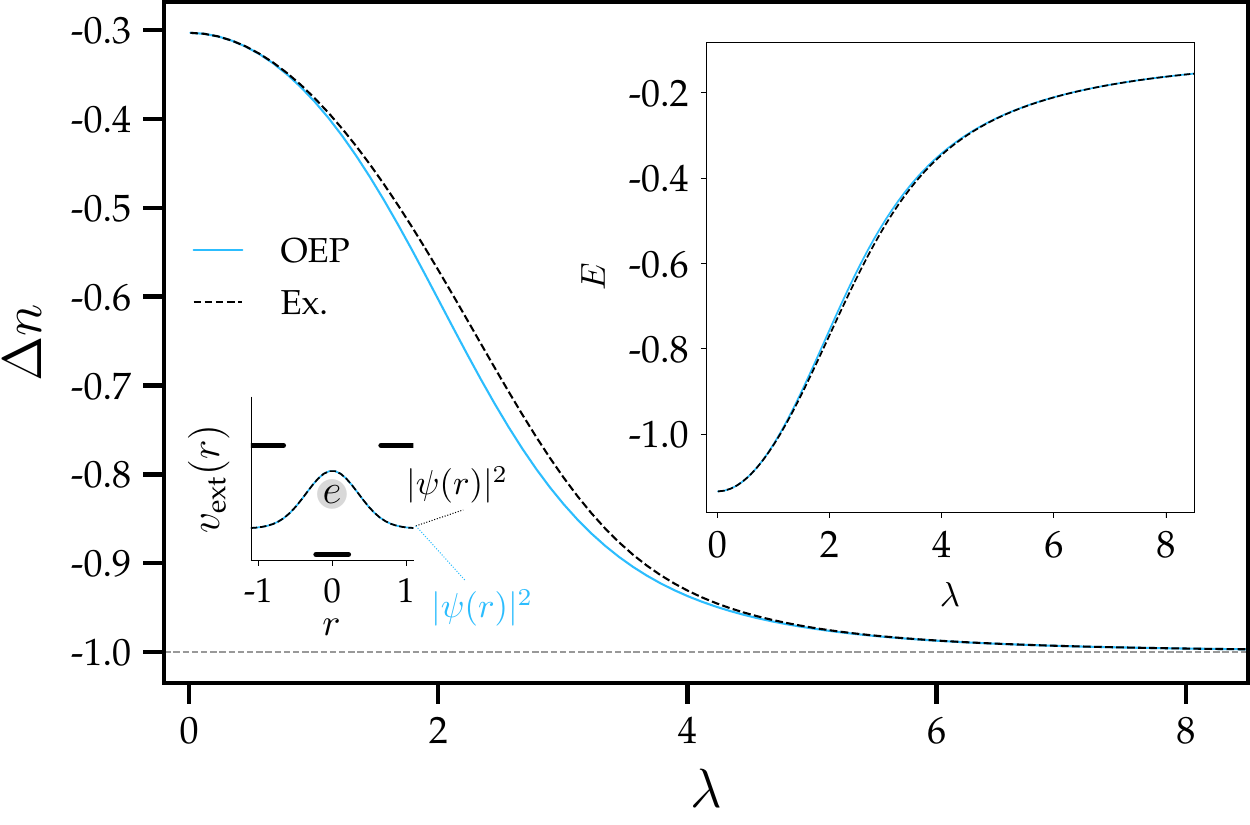}}~~~\subfigure{\label{fig:2}\includegraphics[width=0.49\textwidth]{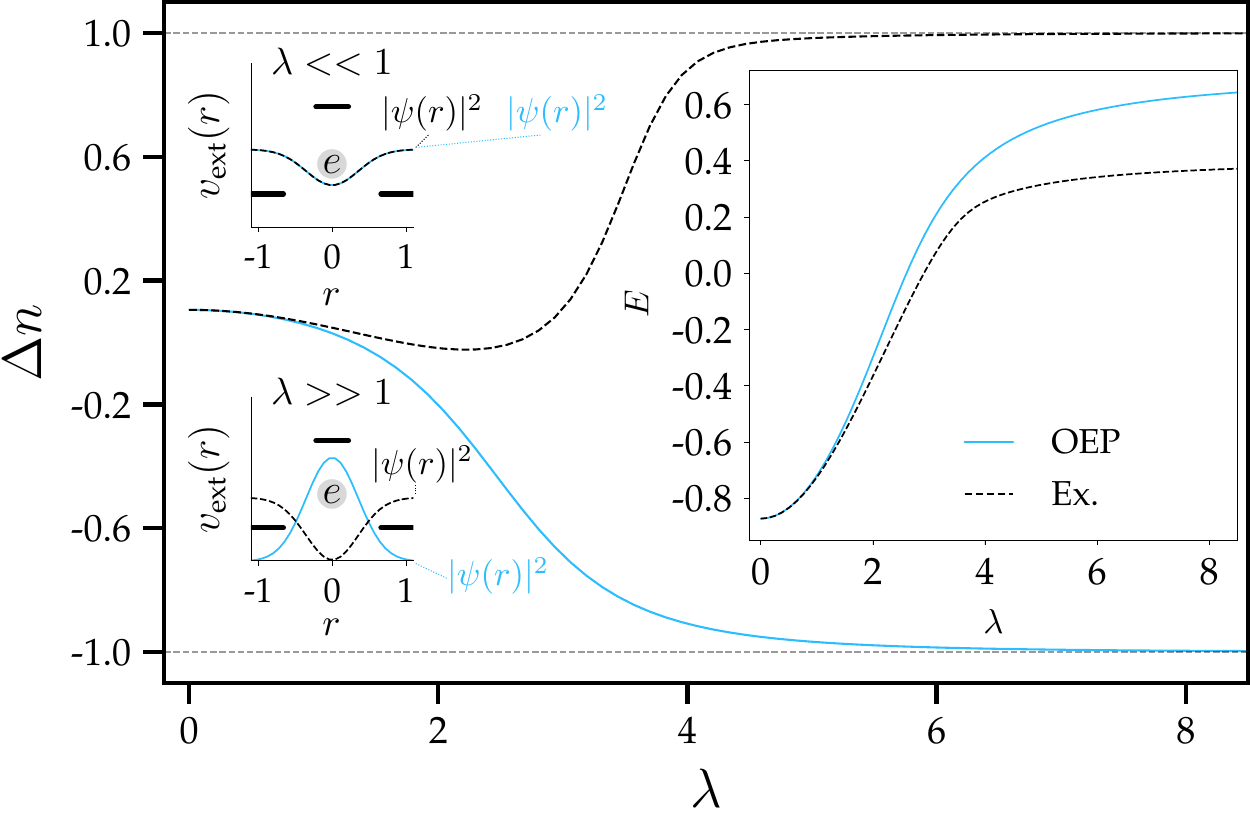}}
\caption{The density disbalance $\Delta n$ and the ground state energy $E_0$ as functions of the coupling constant $\lambda_{\textup{c}}$ obtained from the OEP-QEDFT and from the exact diagonalization for two configurations of the on-site external potentials, $v_{\text{ext}}=\{0.3,-0.6,0.3\}$ (left) and $v_{\text{ext}}=\{-0.1,0.2,0.1\}$ (right). The photon frequency in both cases is $\omega=1.0$.}
\label{fig:none1}
\end{figure*}

Let us rewrite Eqs.~\eqref{eqn:TDOEP} and~\eqref{eqn:s_matrix} for our minimal three-site electron-photon system, and specify all quantities entering these equations.  Since we are interested in analyzing the behavior of the system evolving from its ground state, the KS occupation numbers take the following values, $f_{\textup{g}}=1$, $f_{\textup{e}_1}=0$, and $f_{\textup{e}_2}=0$, where subscripts $\textup{e}_1$ and $\textup{e}_2$ indicate the first and the second excited states, respectively. The vectors of KS orbitals $\{\phi_{i}\}$ read 
\begin{eqnarray} \label{eqn:orbitals}
&&\phi_{\textup{g}}=\begin{pmatrix}
\phi_{\textup{g}}^{\textup{l}}  \\
\phi_{\textup{g}}^{\textup{c}}  \\
\phi_{\textup{g}}^{\textup{r}}  
\end{pmatrix}, \ \phi_{\textup{e}_1}=\begin{pmatrix}
\phi_{\textup{e}_1}^{\textup{l}}  \\
\phi_{\textup{e}_1}^{\textup{c}}  \\
\phi_{\textup{e}_1}^{\textup{r}}  
\end{pmatrix}, \ \phi_{\textup{e}_2}=\begin{pmatrix}
\phi_{\textup{e}_2}^{\textup{l}}  \\
\phi_{\textup{e}_2}^{\textup{c}}  \\
\phi_{\textup{e}_2}^{\textup{r}}  
\end{pmatrix}.
\end{eqnarray}
The evolution of each orbital obeys the time-dependent KS equation with the Hamiltonian of Eq.~\eqref{eqn:ks_syst},
\begin{equation}\label{eqn:KS_three_site}
\textup{i}\partial_t\phi_{i}=\hat{H}_{\textup{KS}}\phi_{i}.
\end{equation}
The dipole matrix elements take the following form:
\begin{equation}
d^{\alpha}_{i,j}=\lambda_{\textup{c}}\big[-(\phi_{i}^{\textup{l}})^*
\phi_{j}^{\textup{l}}
+(\phi_{i}^{\textup{r}})^*
\phi_{j}^{\textup{r}}\big], \quad i,j=\textup{g},\textup{e}_1,\textup{e}_2.
\end{equation}
In the chosen gauge ($v_{\textup{xc}}^{\textup{c}}=-v_{\textup{xc}}^{\textup{l}}-v_{\textup{xc}}^{\textup{r}}$), the TDOEP equations~\eqref{eqn:TDOEP} and~\eqref{eqn:s_matrix} can be written as the following system of two integral equations,
\begin{equation}\label{eqn:oep_three_site_specific_1}
\begin{split}
  &\int_{0}^{t}dt_1\Big[K_{11}(t,t_1)v_\textup{xc}^\textup{l}(t_1)+K_{12}(t,t_1)v_\textup{xc}^\textup{r}(t_1)\Big]=g_1(t,\omega),\quad \ \ \\
  &\int_{0}^{t}dt_1\Big[K_{21}(t,t_1)v_\textup{xc}^\textup{l}(t_1)+K_{22}(t,t_1)v_\textup{xc}^\textup{r}(t_1)\Big]=g_2(t,\omega),\quad \ \ 
\end{split}
\end{equation}
\noindent where $g_i$ and $K_{ij}$ are functions of orbitals $\phi_{m}^{n}$. Thus, in order to determine the evolution of the KS potential $v_{\textup{s}}$, the KS orbitals, and as a consequence the on-site densities,  Eqs.~\eqref{eqn:KS_three_site} and~\eqref{eqn:oep_three_site_specific_1} must be solved self-consistently -- step by step. We note that the photon parameters appear only in the functions $g_1$ and $g_2$. The difference between the single-mode and multiple-mode (modeling the dissipative environment) description is that the functions $g_1$ and $g_2$ involve an additional frequency integration with a chosen spectral density,
\begin{equation}
    g_i(t,\omega)\rightarrow \sum_{\alpha}\rho_{}(\omega_{\alpha})g_i(t,\omega_{\alpha}).   
\end{equation}
From the computational point of view, a standard way of solving the system of Volterra integral equations of the first kind, Eqs.~\eqref{eqn:oep_three_site_specific_1}, is to discretize time ($t_n=t_0+n\Delta_t$, where we chose constant step $\Delta_t$) according to the trapezoidal rule. It is the simplest discretization scheme among linear multistep methods presented in Ref.~\cite{1981Andrade} in case when kernels, $K_{ij}(t_1,t_2)$, are identically zero at $t_1=t_2$. The first step of the procedure reads as:
\begin{eqnarray}\label{eqn:first_step}
&&v_{\textup{xc}}^{\textup{l}}(t_1)K_{11}(t_1,t_0)+v_{\textup{xc}}^{\textup{r}}(t_1)K_{12}(t_1,t_0)=\dfrac{2}{\Delta_t}g_1(t_1,\omega),\qquad\\
&&v_{\textup{xc}}^{\textup{l}}(t_1)K_{11}(t_1,t_0)+v_{\textup{xc}}^{\textup{r}}(t_1)K_{12}(t_1,t_0)=\dfrac{2}{\Delta_t}g_2(t_1,\omega),\qquad
\end{eqnarray}
where for kernels $K_{ij}$ the orbitals at $t=t_0$ and $t=t_1$ are used. Having obtained values of potentials at $t_1$ one can propagate KS orbitals by one time step and find $K_{ij}(t_2,t_1)$. Then, the potentials $v_{\textup{xc}}^{j}$ are calculated as,
\begin{eqnarray}
&&v_{\textup{xc}}^{\textup{l}}(t_n)K_{11}(t_n,t_{n-1})+v_{\textup{xc}}^{\textup{r}}(t_n)K_{12}(t_n,t_{n-1})\nonumber\\
&&=\dfrac{1}{\Delta_t}g_1(t_1,\omega)-\dfrac{1}{2}\Big[v_{\textup{xc}}^{\textup{l}}(t_1)K_{11}(t_n,t_{0})+v_{\textup{xc}}^{\textup{r}}(t_1)K_{12}(t_n,t_{0})\Big]\nonumber\\
&&-\sum\limits_{j=2}^{n-1}\Big[v_{\textup{xc}}^{\textup{l}}(t_j)K_{11}(t_n,t_{j-1})+v_{\textup{xc}}^{\textup{r}}(t_j)K_{12}(t_n,t_{j-1})\Big],\\
&&v_{\textup{xc}}^{\textup{l}}(t_n)K_{21}(t_n,t_{n-1})+v_{\textup{xc}}^{\textup{r}}(t_n)K_{22}(t_n,t_{n-1})\nonumber\\
&&=\dfrac{1}{\Delta_t}g_2(t_1,\omega)-\dfrac{1}{2}\Big[v_{\textup{xc}}^{\textup{l}}(t_1)K_{21}(t_n,t_{0})+v_{\textup{xc}}^{\textup{r}}(t_1)K_{22}(t_n,t_{0})\Big]\nonumber\\
&&-\sum\limits_{j=2}^{n-1}\Big[v_{\textup{xc}}^{\textup{l}}(t_j)K_{21}(t_n,t_{j-1})+v_{\textup{xc}}^{\textup{r}}(t_j)K_{22}(t_n,t_{j-1})\Big].
\end{eqnarray}
In the next section, these equations together with Eq.~\eqref{eqn:KS_three_site} will be propagated for various initial conditions and external potentials.

\section{Results and discussion}
 
\subsection{Performance of OEP in the absence of dissipation: Single-mode cavity} 

Before proceeding with the time-dependent problem, we analyze first a ground state of the system in the presence of a reflection symmetric external potential. Apparently, the KS potential is also symmetric and in our gauge we parametrize it as follows, $v_{\textup{s}}^{\textup{l}}=v_{\textup{s}}^{\textup{r}}=v_{\text{s}}$, and $v_{\textup{s}}^{\textup{c}}=-2v_{\text{s}}$.

The eigenvalues and KS orbitals (up to normalization) for the static KS problem $\hat{H}_{\textup{KS}}\phi_{i}=\varepsilon_{i}\phi_{i}$ are as follows,
\begin{eqnarray}
&&\varepsilon_{\textup{g}}=-\dfrac{v_{\textup{s}}+W}{2}, \ \ \phi_{\textup{g}}=\Bigg[1,\
\dfrac{3v_{\textup{s}}+W}{2T},\ 1\Bigg],\\  
&&\varepsilon_{\textup{e}_2}=\dfrac{W-v_{\textup{s}}}{2}, \ \ \phi_{\textup{e}_2}=\Bigg[1,\
\dfrac{3v_{\textup{s}}-W}{2T},\ 1\Bigg],\\ 
&&\varepsilon_{\textup{e}_1}=v_{\textup{s}}, \ \ \phi_{\textup{e}_1}=\Bigg[-1,\ 0, \ 1\Bigg],
\label{e1}
\end{eqnarray}
where $W=\sqrt{8T^2+9v_{\textup{s}}^2}$. The difference $\Delta n$ of the side and central densities, Eq.~\eqref{eqn:delta_n}, is equal to $-3 v_{\textup{s}}/W$. In the static case, the OEP equation \eqref{eqn:TDOEP} for the potential $v_{\rm x}$ reduce an algebraic transcendental equation, 
\begin{equation}\label{eqn:oep_st_potential}
v_{\text{xc}}=v_{\text{s}}-v_{\textup{ext}}=\dfrac{\lambda^2\big((3v_{\text{s}}+W)^2+6\,\omega v_{\text{s}}\big)}{6(2\omega+3 v_{\text{s}} +W)^2}.
\end{equation}
This equation can also be obtained by minimizing the ground state energy of the total electron-photon system,
\begin{equation}
    E_0=\dfrac{1}{2}\omega-\braket{\phi_{\textup{g}}|\hat{T}|\phi_{\textup{g}}}+\braket{\phi_{\textup{g}}|\hat{V}_{\textup{ext}}|\phi_{\textup{g}}}+E_\textup{xc},
\end{equation}
where $E_{\textup{xc}}$ is the Lamb shift energy determined by the diagram on Fig.~\ref{fig:diag}(a), which explicitly reads as,
\begin{equation}
    E_{\textup{xc}}=\dfrac{\lambda^2 (W^2-9 v_{\textup{s}}^2)}{4 W (2 \omega + 3 v_{\textup{s}} + W)}.
\end{equation}
\begin{figure*}[t!]
\subfigure{\label{fig:3}\includegraphics[width=0.49\textwidth]{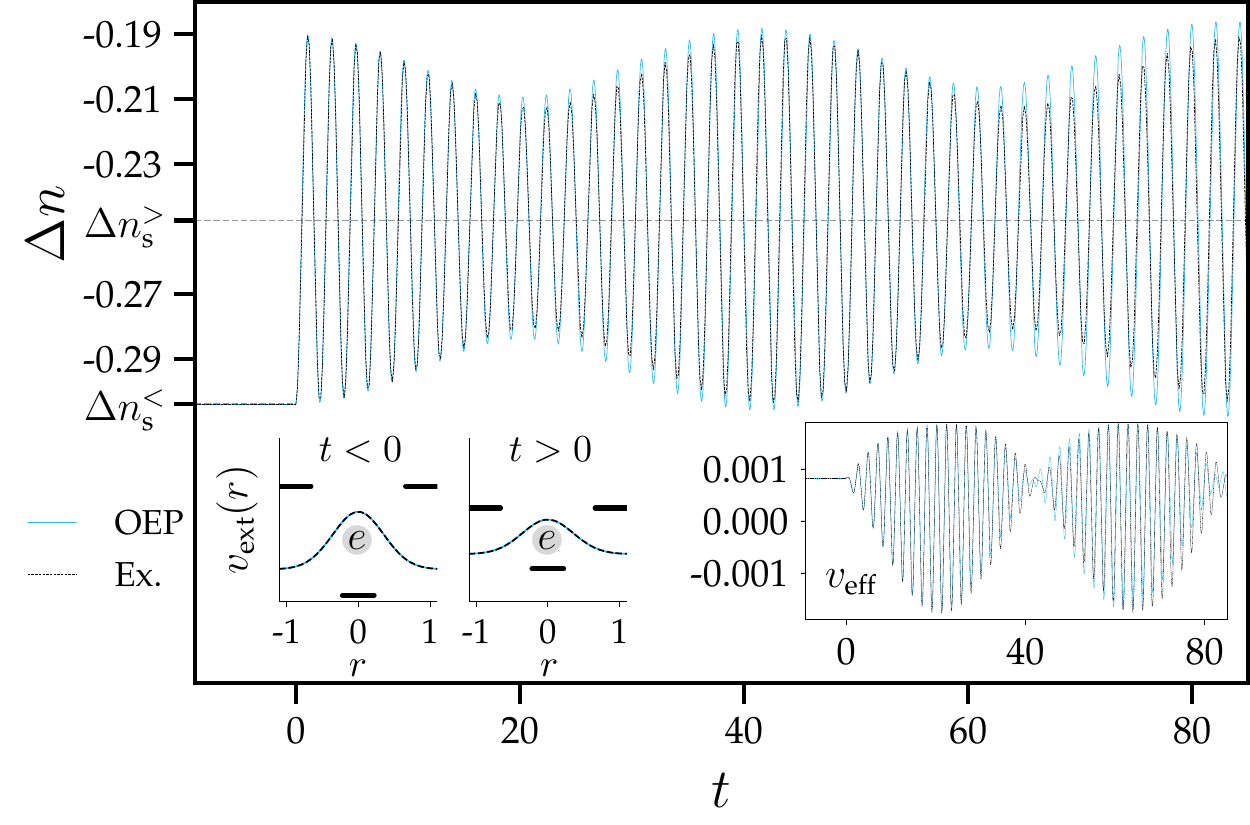}}~~~\subfigure{\label{fig:4}\includegraphics[width=0.49\textwidth]{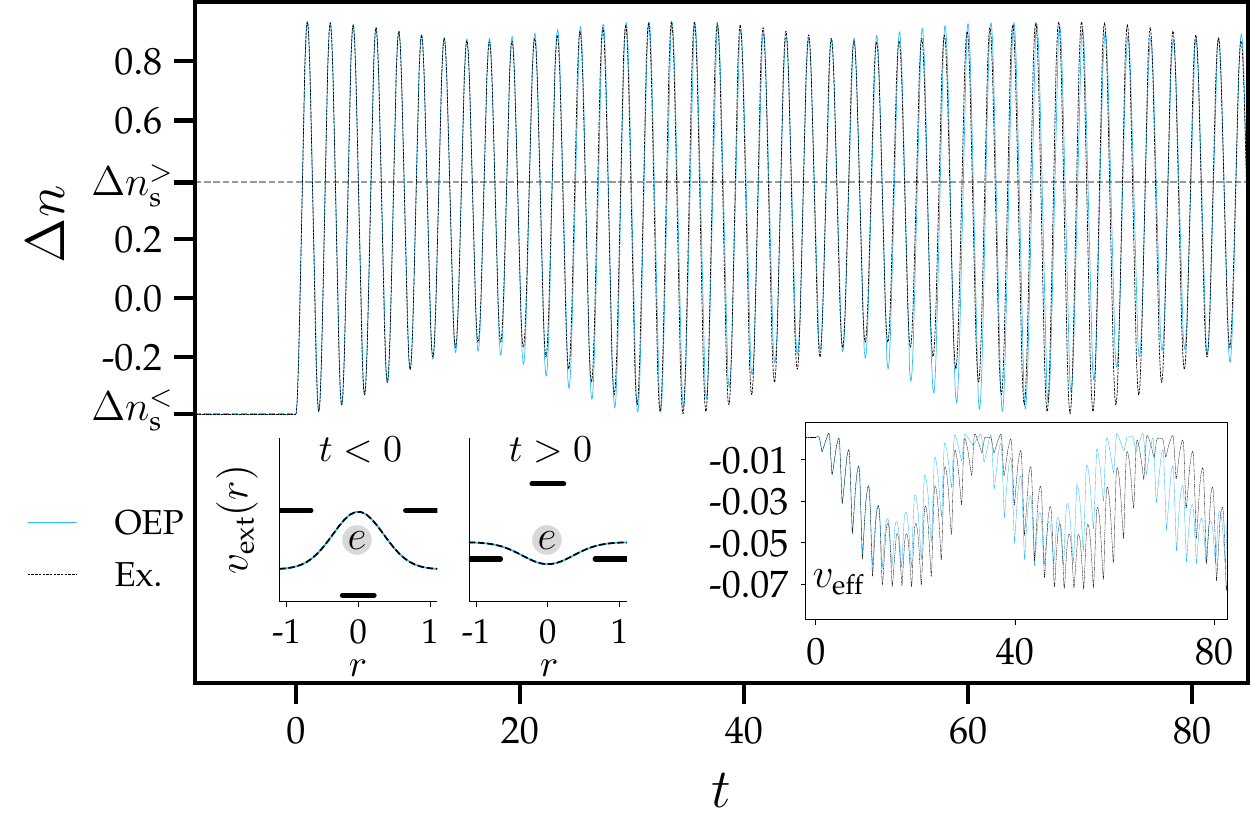}}
\caption{Dynamics of $\Delta n$ and the xc potential $v_{\textup{xc}}$ in a single-mode cavity generated by switching the external potential at $t=0$ from  $v_{\text{ext}}^{t<0}=\{0.3,-0.6,0.3\}$ to $v_{\text{ext}}^{t>0}=0.8\,v_{\text{ext}}^{t<0}$ (left) and to $v_{\text{ext}}^{t>0}=-\,v_{\text{ext}}^{t<0}$ (right). In both cases the coupling constant $\lambda=0.1$, while the photon frequency $\omega$ is $1.0$ and $2.0$ for the left and right panels, respectively. OEP-QEDFT and exact results are shown by blue and black lines, respectively.}
\label{fig:none2}
\end{figure*}

In all numerical calculations below we choose the hopping rate $T$ as a unit of energy, that is, we set $T=1$. For definiteness, we analyze $\Delta n$ and $E_0$ for two mutually inverted configurations of the external potential, $v_{\textup{ext}}=(0.3,-0.6,0.3)$ and $v_{\textup{ext}}=(-0.1,0.2,-0.1)$, which correspond to a potential well or a hump located at the central cite. In Figs.~\ref{fig:none1} we show the calculated OEP density and total energy as functions of the coupling strength $\lambda$ and compare them with the results obtained by the exact diagonalization of the Hamiltonian~\eqref{eqn:latt_ham} in a properly truncated Fock space. For the first configuration, shown in Fig.~\ref{fig:none1}(left), OEP works quite well practically for any $\lambda$, becoming essentially exact in the weak and ultra strong coupling regimes. This behavior is not surprising, and very similar to the picture observed for a two-site model~\cite{Pellegrini2015PRL}. For a very strong coupling both OEP-QEDFT and the exact solution predict localization of the electron at the central site (with lower potential), which physically reflects the formation of a small radius polariton and the corresponding suppression of the tunneling. A very different picture is observed if we invert the external potential and take $v_{\textup{ext}}=(-0.1,0.2,-0.1)$, see Fig.~\ref{fig:none1}(right). For relatively weak couplings with $\lambda\le 1$ OEP still shows good results, which is expected for a perturbative construction, but it fails dramatically in the ultra strong coupling regime. OEP still localizes the electron at the center, while in the exact solution it is trapped on the side sites with lower potential, in agreement with the physical picture of a small radius polariton and suppression of the tunneling. The reason for this failure at strong couplings is that the growth of the Lamb shift energy $E_{\textup{xc}}$ in the OEP functional can only be suppressed by minimizing the dipole matrix element between the ground and the first excited KS orbitals. Since the first excited orbital is fixed by the symmetry, see Eq.~\eqref{e1}, the overlap is minimized by localizing the ground KS state, and thus the density, at the central site. We emphasize that the detected problems of QED-OEP show up only in regime of ultra strong coupling corresponding to small radius polaritons and can hardly be realized in practice.

We therefore conclude that for sufficiently small coupling constants $\lambda\le 1$ OEP approximation produces good results for ground state properties. Expecting a similar behavior for dynamics, in the following we will never consider coupling constants exceeding $\lambda=0.4$.

\begin{figure*}[t!]
\subfigure{\label{fig:5}\includegraphics[width=0.49\textwidth]{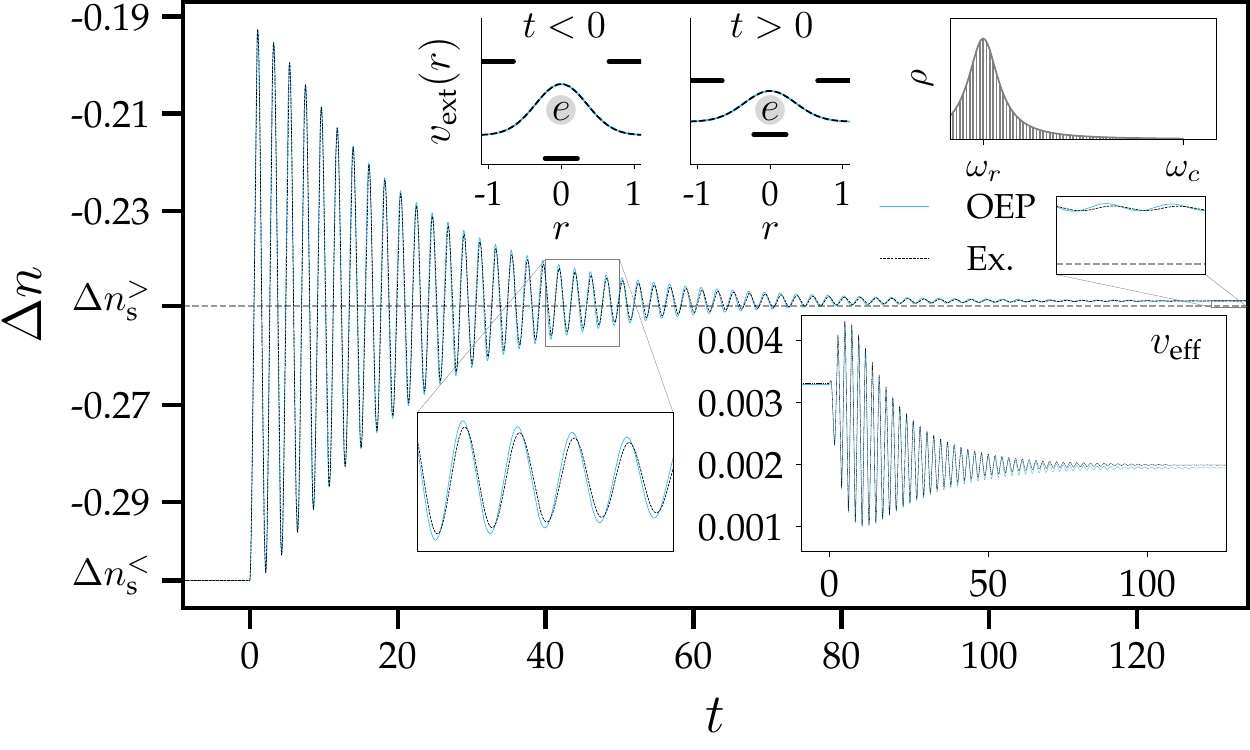}}~~~\subfigure{\label{fig:6}\includegraphics[width=0.49\textwidth]{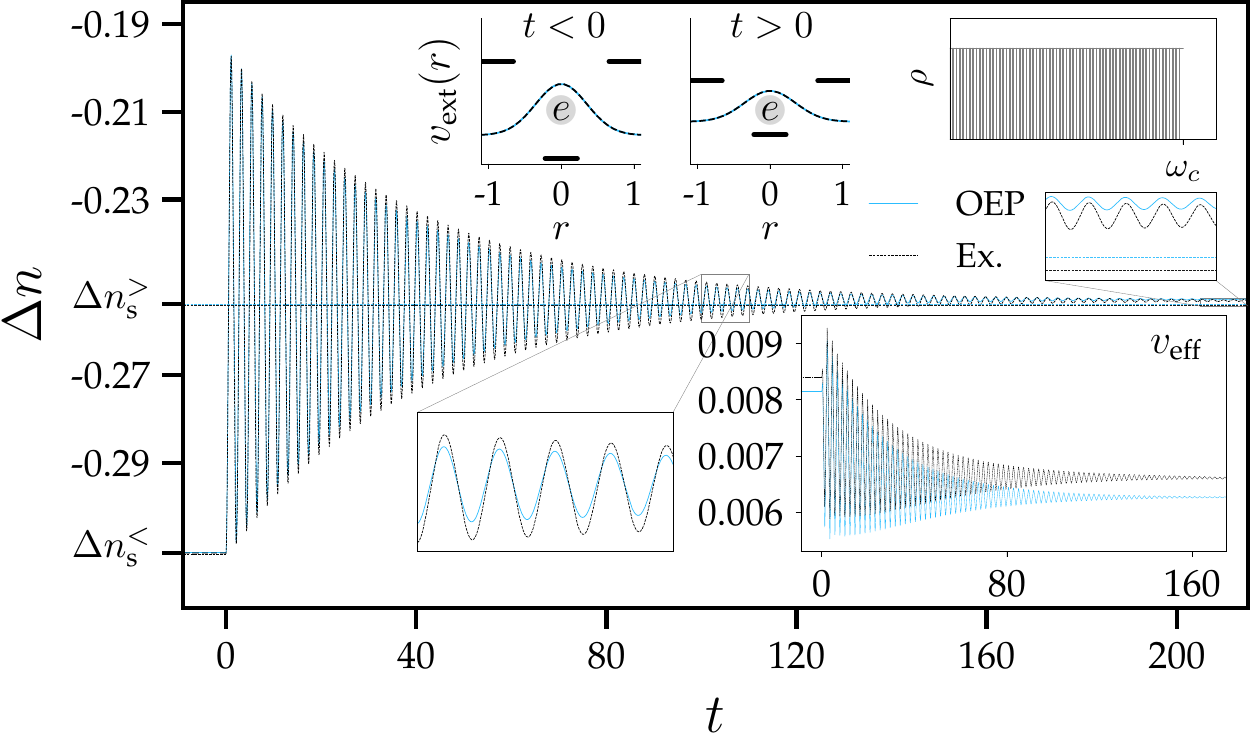}}
\caption{Relaxation dynamics of $\Delta n$ and the corresponding xc potential $v_{\textup{xc}}$ obtained from OEP-QEDFT (blue solid lines) and from the exact solution (black solid lines). The dynamics is generated as on Fig.~\ref{fig:none2} (left) by switching the potential from $v_{\text{ext}}^{t<0}=\{0.3,-0.6,0.3\}$ to $v_{\text{ext}}^{t>0}=0.8\,v_{\text{ext}}^{t<0}$. 
Left panel: Dissipation in a lossy cavity with one photon peak of Lorentzian shape, Eq.~\eqref{eqn:lor_prof}.
Right panel: Dissipation for Ohmic bath -- the flat spectral density of Eq.~\eqref{eqn:unif_prof}. 
Adopted parameters: $\lambda$ is $0.2$ (left) and $0.4$ (right), the resonance frequency $\omega_\textup{r}=2.0$ (relevant for the left panel only), cutoff $\omega_\textup{c}$ is $7.0$ (left) and $6.0$ (right), loss rate $\gamma=0.3$ (relevant for the left panel only), number of photon modes $N_{\gamma}$ is $1050$ (left) and $700$ (right) .}
\label{fig:none3}
\end{figure*}

Let us now analyze dynamics of the system in the singe-mode regime. Specifically we consider dynamics generated by switching the external potential. In the first example, we prepare the system in the ground state in the external potential $v_{\text{ext}}^{t<0}=\{0.3,-0.6,0.3\}$, which at $t=0$ is suddenly rescaled by the factor $0.8$, that is $v_{\text{ext}}^{t>0}=0.8\,v_{\text{ext}}^{t<0}$. The generated dynamics presented in Fig.~\ref{fig:none2}(left) demonstrates a good agreement of QED-OEP results with the exact solution of the electron-photon problem. The function $\Delta n(t)$ shows typical quantum beats at the frequency of main electronic transition, which in our case is also of the order of the photon frequency. The main qualitative effect of the electron-photon coupling is a periodic modulation of the beats' amplitudes with a smaller frequency that depends on the coupling constant $\lambda$ and can be interpreted as an effective Rabi frequency. The $\lambda$-dependence of the modulation frequency is indeed perfectly fitted with a typical expression for the Rabi frequency of a two-level system interacting with a photon mode, 
\begin{equation}\label{eqn:rabi_freq}
\sim\sqrt{\Delta^2+\lambda^2},    
\end{equation}
where $\Delta=\omega-\omega_{\textup{trans}}$ is the detuning of the photon frequency $\omega$ from the transition frequency $\omega_{\textup{trans}}$.  

In our second example, presented on Fig.~\ref{fig:none2}(right), we start from the same initial state, but generate a more nonlinear dynamics by inverting the potential at $t=0$, which is a much stronger perturbation.  However here we still observe a good qualitative and quantitative performance of QEDFT with OEP potential compared to the direct solution of the electron-photon problem.

Importantly, because of the symmetry, the classical radiation and the corresponding mean-field (radiation reaction) potential are totally absent. All effects of the electron-photon coupling, which are obviously quite significant, should be attributed to quantum xc effects related to the spontaneous radiation channel. Apparently OEP approximation captures these effects quite well. Below we will see that this still holds true in the presence of dissipation, where the importance of xc effects becomes even more pronounced.

\subsection{Quantum dissipation in a lossy cavity}
We now turn to electron dynamics in the case of coupling to a dissipative environment represented by a continuum of photon modes in a lossy cavity. 
In this section we again analyze previously considered settings with dynamics generated by switching the external potential. However, now we add coupling to a photon continuum with two types of spectral densities described in Sec.~II~B. First, we consider a lossy cavity with a single photon peak broadened according to Lorentzian distribution of Eq.~\eqref{eqn:lor_prof}. As a second example we study dynamics in the presence of the Ohmic bath modeled by photon modes with a flat spectral density, Eq.~\eqref{eqn:unif_prof}. 

\begin{figure*}[t!]
\subfigure{\label{fig:7}\includegraphics[width=0.49\textwidth]{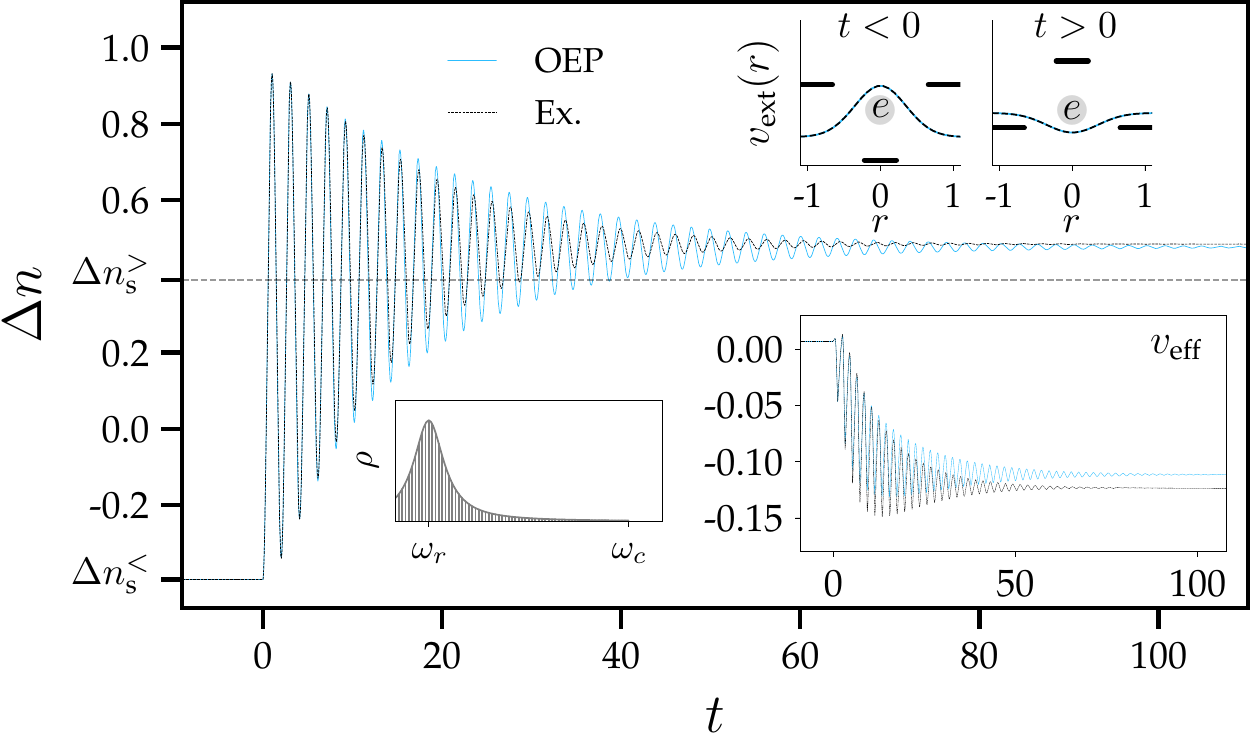}}~~~\subfigure{\label{fig:8}\includegraphics[width=0.49\textwidth]{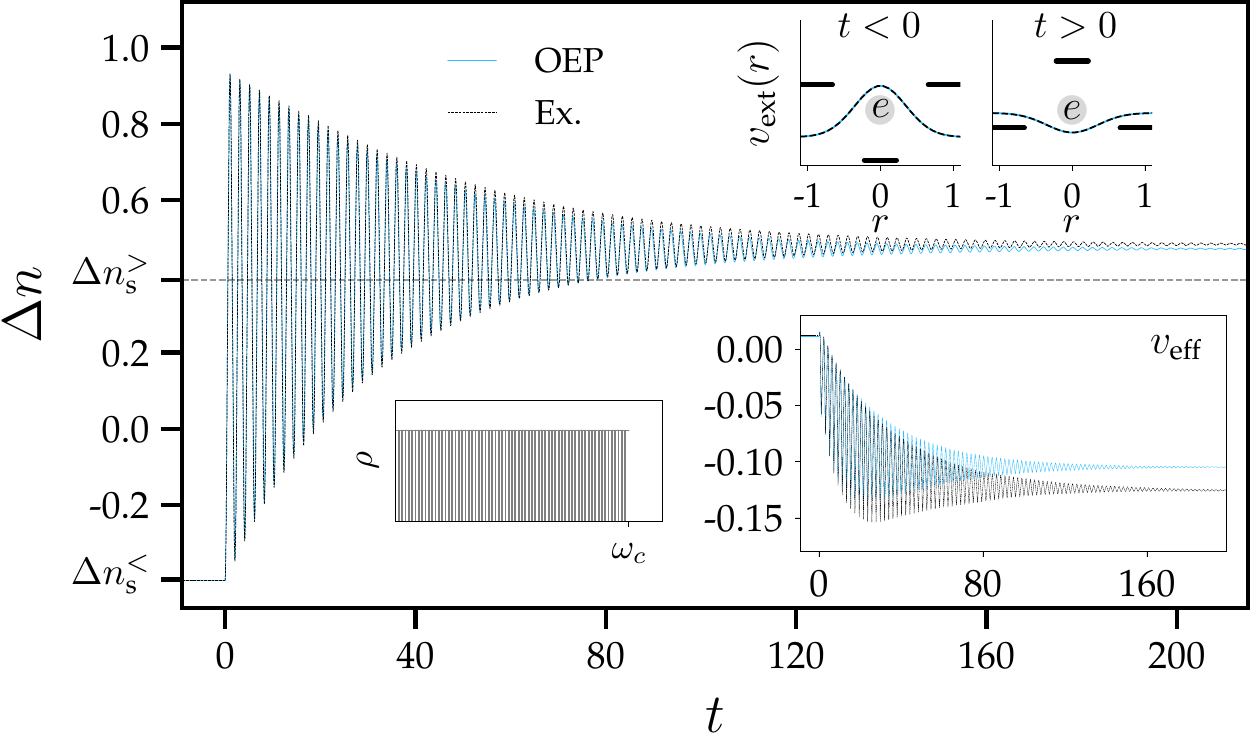}}
\caption{Relaxation dynamics of $\Delta n$ and the corresponding xc potential $v_{\textup{xc}}$ obtained from OEP-QEDFT (blue solid lines) and from the exact solution (black solid lines). The dynamics is generated by suddenly inverting the external potential at $t=0$, from  $v_{\text{ext}}^{t<0}=\{0.4,-0.8,0.4\}$ to $v_{\text{ext}}^{t>0}=-v_{\text{ext}}^{t<0}$.  
Left panel: Dissipation in a lossy cavity with one photon peak of the Lorentzian shape, Eq.~\eqref{eqn:lor_prof}.
Right panel: Dissipation for Ohmic bath -- the flat spectral density of Eq.~\eqref{eqn:unif_prof}.
Adopted parameters: $\lambda$ is $0.3$ (left) and $0.4$ (right), the resonance frequency $\omega_\textup{r}=2.0$ (relevant for the left panel only), cutoff $\omega_\textup{c}$ is $7.0$ (left) and $6.0$ (right), loss rate $\gamma=0.5$ (relevant for the left panel only), number of photon modes $N_{\gamma}$ is $500$ (left) and $700$ (right) .}
\label{fig:none4}
\end{figure*}

The evolution of the density $\Delta n(t)$ and the xc potential $v_{\textup{xc}}(t)$ for two different types of potential switching is presented in Figs.~\ref{fig:none3} and \ref{fig:none4}. In all our examples the system shows relaxation dynamics in which, after a sudden switch, the density distribution adapts to a new shape of the potential. In the case of the single Lorentzian peak in the spectral density one could expect, by analogy with Fig.~\ref{fig:none2}, to see some decaying Rabi oscillations. However for our parameters the characteristic relaxation time is shorter than one Rabi cycle. Therefore, the Rabi oscillations are overdamped. In fact, in the left panels in Figs.~\ref{fig:none3} and \ref{fig:none4} we see a qualitatively similar, practically exponential decay of quantum beats both for the Lorentzian and for the flat distribution of photon modes.  

When the electron density relaxes when adopting to the new potential, the energy of the electron subsystem is transferred to the cavity photons. Due to the inversion symmetry the dipole moment is always zero and the classical radiation is completely suppressed. This means that at the mean-field level we would get persistent beats and no relaxation/decay.  The relaxation we see in Figs.~\ref{fig:none3} and \ref{fig:none4} is a purely quantum effect of a spontaneous incoherent radiation and the corresponding energy transfer from the electron to the cavity photons. In formalism of QEDFT this is a purely xc effect encoded in the xc potential. Our results clearly show that xc corrections to the electron dynamics are huge, and OEP approximation works surprisingly well in all regimes considered in this work.

Apparently, the OEP approximations perfectly captures the main xc effects responsibly for quantum dissipation. However our results also demonstrate some deficiencies of OEP which show up in the long-time asymptotic regime.

Let us first analyse the dynamical regime presented on Fig.~\ref{fig:none3}. It is clear physically that in the long-time limit $\Delta n(t)$ should approach its value $\Delta n^>_{\textup{s}}$ in the new ground state. On Fig.~\ref{fig:none3} the value of $\Delta n^>_{\textup{s}}$ obtained solving the stationary problem is shown by a horizontal dashed line By magnifying the asymptotic region we observe that the OEP result does not converge to the expected ground state value. Surprisingly, OEP reproduces perfectly the "exact" results and therefore both dynamical methods yield identical, but wrong asymptotic densities. At this point we recall that for benchmarking dissipative dynamics we restricted the consideration to only one photon states (see Sec.~III~A). We therefore conclude that OEP almost ideally describes dissipation dynamics dominated one-photon processes. In the present case the deviation from correct asymptotic values is about a tenth of a percent for potentials, which are the most sensitive indicators of the quality, and even less for densities. Therefore the observed error is practically irrelevant. However, in the regime presented in Fig.~\ref{fig:none4}, when dynamics is generated by a stronger perturbation, the deviation of the asymptotic density from the expected new ground state value is much larger and visible without any magnification. The OEP is still in a good agreement with the exact one-photon calculations. This indicates that the error is apparently due to missing two-photon processes, which, of course, are important in the case of a three-level system. Nonetheless an overall performance of OEP-QEDFT in capturing quantum dissipation is still quite reasonable.

It should be added here that from the computational point of view, especially in considering dissipation processes, the OEP approach turns out to be extremely beneficial in terms efficiency and calculation time, even if in the direct solution of the electron-photon Schr\"odinger equation the Fock space is truncated to one-photon states. 

\section{Conclusion}

In conclusion, we demonstrate the possibility of describing quantum dissipation in the framework of QEDFT with xc potential approximated within the OEP formalism.  This opens a way for the first principle modeling of non-relativistic electron systems interacting with cavity photons of realistic lossy cavities, as well as with other types of Caldeira-Leggett dissipative environments relevant in condensed matter and chemical physics. 

This work should be considered as a proof of principle for the applicability of QEDFT in general, and OEP in particular to quantum dissipative systems. Using a minimal three-site model we showed that for moderate values of the coupling constant the lowest order conserving OEP performs very good qualitatively in different regimes, and is in excellent quantitative agreement with the exact solution provided the dissipation is dominated by one-photon processes. By a special choice of inversion symmetric external potential we completely suppress the classical radiation, and thus prove unambiguously that OEP captures the main quantum features of spontaneous radiation that significantly modifies dynamics of relevant observables.  

In principle the present dissipative version of the OEP-QEDFT can be directly employed for the quantitative  modelling of the cavity assisted photocatalysis and more generally, polaritonic chemistry experiments with realistic lossy cavities. Unfortunately, the QED-OEP suffers from the same conceptual problems as the standard OEP \cite{KumKro2008}, being quite expensive computationally. One of the main problems is the necessity to propagate all, but not only occupied, KS orbitals. In this respect the numerical efficiency can probably be improved using the Sternheimer formalism, as it has been done recently for the ground state QED-OEP \cite{Flick2018}. It is however not clear for the moment how to extend this to the time-dependent setting. Obviously the ideal way of making QEDFT practical is to develop local or semilocal functionals of LDA, GGA, or, possibly, Vignale-Kohn \cite{VigUllCon1997} type. The latter framework looks especially promising for capturing dissipative xc effects. One practical outcomes of the present work is that for development of new more efficient dissipative functionals for QEDFT, the QED-OEP can serve as a trustable benchmark in those cases when the exact solution is not possible.       

\section*{Acknowledgement}
We are grateful to D. Gulevich for insightful discussion and valuable comments. The work was supported by Russian Science Foundation (Project No. 20-12-00224). I.V.T. acknowledges support by Grupos Consolidados UPV/EHU del
Gobierno Vasco (Grant No. IT1249-19) and by Spanish MICINN (Project No. PID2020-112811GB-I00).

\bibliography{cavityQED.bib}

\end{document}